\documentclass{aa}
\usepackage[dvips]{graphics}
\begin{document}

   \thesaurus{03(11.03.4; 11.04.1; 11.05.1; 11.06.2; 11.19.2; 12.04.3)}
   \title{Structure, mass and distance of the Virgo cluster from a Tolman-Bondi
          model}
   \author{Pascal Fouqu\'e
          \inst{1,2}
          \and Jos\'e M.\ Solanes
          \inst{3}
          \and Teresa Sanchis
          \inst{4}
          \and Chantal Balkowski
          \inst{5}
          }
   \offprints{P. Fouqu\'e}
   \institute{Observatoire de Paris-Meudon DESPA, F-92195 Meudon CEDEX, France
        \and European Southern Observatory, Casilla 19001, Santiago 19, Chile\\
             email: pfouque@eso.org
        \and Departament d'Enginyeria Inform\`atica i Matem\`atiques,
             Escola T\`ecnica Superior d'Enginyeria, 
             Universitat Rovira i Virgili,
             Carretera de Salou, s/n; E--43006~Tarragona, Spain\\
             email: jsolanes@etse.urv.es
        \and Departament d'Astronomia i Meteorologia,
             Facultat de F\'{\i}siques, Universitat de Barcelona
             C/ Mart\'{\i} i Franqu\'es 1; E--08028~Barcelona, Spain\\
             email: tsanchis@am.ub.es
        \and Observatoire de Paris-Meudon DAEC, F-92195 Meudon CEDEX, France\\
             email: Chantal.Balkowski@obspm.fr
             }
   \date{Received / Accepted }
   \authorrunning{P. Fouqu\'e et al.}
   \titlerunning{Structure, mass and distance of the Virgo cluster}
   \maketitle
   \begin{abstract}

   We have applied a relativistic Tolman-Bondi model of the Virgo cluster to a 
sample of 183 galaxies with measured distances within a radius of 8 degrees 
from M~87. We find that the sample is significantly contaminated by background 
galaxies which lead to too large a cluster mean distance if not excluded. The 
Tolman-Bondi model predictions, together with the HI deficiency of spiral 
galaxies, allows one to identify these background galaxies. One such galaxy is 
clearly identified among the 6 calibrating galaxies with Cepheid distances. As 
the Tolman-Bondi model predicts the expected distance ratio to the Virgo 
distance, this galaxy can still be used to estimate the Virgo distance, and the
average value over the 6 galaxies is $15.4 \pm 0.5$ Mpc.

   Well-known background groups of galaxies are clearly recovered, together 
with filaments of galaxies which link these groups to the main cluster, and are
falling into it. No foreground galaxy is clearly detected in our sample.
Applying the B-band Tully-Fisher method to a sample of 51 true members of 
the Virgo cluster according to our classification gives a cluster distance of
$18.0 \pm 1.2$ Mpc, larger than the mean Cepheid distance.

   Finally, the same model is used to estimate the Virgo cluster mass, which is
$M = 1.2\ 10^{15} M_{\sun}$ within 8 degrees from the cluster center (2.2~Mpc
radius), and amounts to 1.7 virial mass.

      \keywords{Galaxies: clusters: individual: Virgo
                 -- Galaxies: distances and redshifts
                 -- Galaxies: elliptical and lenticular
                 -- Galaxies: fundamental parameters
                 -- Galaxies: spiral
                 -- Cosmology: distance scale
               }
   \end{abstract}

\section{Introduction}

   In 1990, Fouqu\'e et al. (\cite{fou90}) derived an unbiased distance to the 
Virgo cluster, based on a complete sample of 178 spiral galaxies, using the 
$B$-band Tully-Fisher relation (Tully \& Fisher \cite {tul77}). A previous 
similar study based on 110 spiral galaxies (Sa - Sm) was published by 
Kraan-Korteweg et al. (\cite{kra88}). Soon after, Teerikorpi et al. 
(\cite{tee92}, hereafter T92) suggested that this distance determination may 
have been contaminated by the inclusion of background galaxies into the 
complete sample, although we a priori excluded galaxies generally attributed to
the background M group and W cloud.

   The study of the structure of the Virgo cluster starts with de Vaucouleurs
(\cite{dev61}), who identifies the southern extension (Virgo cloud X), and the 
wing (Virgo cloud W). This study is extended in de Vaucouleurs \& de 
Vaucouleurs (\cite{dev73}) who separate the Virgo I cluster (Virgo E, S and S')
from the Virgo II cloud complex (composed of Virgo V, X and Y) and the 
background W cloud (composed of Virgo Wa, Wb and W' sub-groups). Paturel 
(\cite{pat79}) applies taxonomy to disentangle these various components. Then, 
Tully (\cite{tul82}) discusses the separation between the Virgo cluster and its
southern extension in the frame of the Local Supercluster. Later, Ftaclas et 
al. (\cite{fta84}) identify the M group and the N group. Pierce \& Tully
(\cite{pie88}) use the Tully \& Shaya infall model (Tully \& Shaya 
\cite{tul84}) to study the velocity - distance diagram of bright Virgo spiral 
galaxies. Binggeli et al. (\cite{bin93}) clarify the situation by listing
galaxies from the VCC catalogue (\cite{bin85}, hereafter VCC) belonging to each
of these clouds (W, W', M and southern extension). T92 list several regions in 
the velocity - distance diagram: A, B, C1, C2, and D. Yasuda et al. 
(\cite{yas97}) also study the 3-D structure of the cluster. Finally, 
Gavazzi et al. (\cite{gav99}, hereafter G99) use distance determinations to 
identify new groups, named B, E, N, and S.

   Please note that a clarification of the nomenclature is highly desirable, as
the same letters refer to totally different groups: A can mean the main Virgo 
cluster or refer to galaxies with high velocities in front of Virgo and falling
into the cluster (T92); B can mean the concentration around M~49, or a
background group in the same region (G99), or even a foreground expanding 
component (T92); E can refer to the elliptical component of the Virgo cluster 
or an eastern group (G99); N can be a group identified by Ftaclas et al. 
(\cite{fta84}) or a northern group in G99; S refers to the spiral component 
of the Virgo cluster or a southern group in G99; X can mean that the galaxy 
lies within the X-ray contours (Federspiel et al. \cite{fed98}, hereafter F98) 
or belongs to the Virgo X cloud (de Vaucouleurs \cite{dev61})!

   To investigate the structure of the Virgo cluster and determine the mean 
distance of the main cluster, we have used the Tolman-Bondi model of the 
cluster defined in Ekholm et al. (\cite{ekh99}, hereafter E99) in the spirit 
of T92, for spiral galaxies whose distances were given by the Tully-Fisher 
relation in $B$-band by Ekholm et al. (\cite{ekh00}, hereafter E00) or in 
$H$-band by G99, but adding the information given by the HI deficiency of 
Virgo spiral galaxies. We also used the same model for early-type galaxies 
whose distances were determined by the fundamental plane method in G99, or 
the recent Tonry's compilation of Surface Brightness Fluctuations distances 
(Tonry et al. \cite{ton01}, hereafter T01).

   Section~2 defines the observable parameters we will use for this study.
Section~3 introduces the Tolman-Bondi model and lists its adopted parameters.
Section~4 investigates how the discrepant Cepheid distance to NGC~4639
is well explained by our model, and computes the average Cepheid distance to 
Virgo. Section~5 applies our model to the determination of the Virgo cluster 
mass, and compares it to other determinations. Section~6 investigates the 
structure of the Virgo cluster and lists galaxies which do not belong to the 
main cluster according to our model. Section~7 compares the average distances 
to Virgo adopted in the various references we used to what our model predicts. 
Finally, Sect.~8 summarizes the main results of this paper.

\section{Observable parameters}

   The selected sample has been extracted from the LEDA database. It consists 
of 584 galaxies within a radius of 8 degrees (2.25~Mpc at 16~Mpc) from 
M~87 ($\alpha_{2000} = 12.51381$, $\delta_{2000} = 12.3900$) and known
recession velocities smaller than 3000 km/s. Unfortunately, not all these 
galaxies have distance estimates, and we therefore reduced the working sample 
to 125 late-type and 67 early-type galaxies, with distance measurements from 
G99, E00, F98 and T01. Let us describe in more detail some of the 
observable parameters collected for these galaxies.

\subsection{Coordinates and recession velocity}

   These parameters are directly extracted from the LEDA database. Coordinates 
are given in the J2000.0 equinox, and helocientric velocities are corrected to 
the Local Group centroid using the Yahil et al. (\cite{yah77}) formula. A 
histogram of the corrected velocities for the full sample of 584 galaxies is 
presented in Fig.~\ref{histov}. It clearly shows more galaxies on the 
high-velocity side than expected from a gaussian distribution. A similar but 
smaller effect also seems apparent on the low-velocity side. For galaxies
with distance measurements, the same features appear.

\begin{figure}
   \begin{center}
   \scalebox{0.4}{\includegraphics{histov.eps}}
   \end{center}
   \caption{Histogram of the recession velocities, in the Local Group reference
   frame, for the 584 galaxies within 8 degrees from M~87. The dotted histogram
   corresponds to galaxies with distance measurements}
   \label{histov}
\end{figure}

\subsection{Morphological types}

   An accurate estimate of morphological types is important at least for two 
reasons: the HI deficiency parameter (see below) is calculated by comparing
the HI content of a galaxy to the average HI content of an isolated galaxy of 
the same morphological type. Any error in the type therefore translates into a 
corresponding error in the HI deficiency estimate. Similarly, distances 
determined by E00 use the calibration of Theureau et al. (\cite{the97}), 
where the intercept of the Tully-Fisher relation varies with the type. Any 
error in the type changes the galaxy distance value accordingly.

   We used three sources of morphological type determinations: 
de Vaucouleurs et al. (\cite{dev91}, hereafter RC3), the VCC catalogue and 
van den Bergh et al. (\cite{van90}). The quality of RC3 types depends upon its 
source, coded according to Table~3 of Vol.~1 of the catalogue. If the 
morphological type has been measured on large reflector plates (code R), it is 
adopted; if it comes from PSS prints (codes P or U), we adopt an eye average 
with the two other sources. The adopted morphological types differ therefore 
from those given in LEDA and used by E00.

   We adopted a numeric morphological type according to the RC3 coding scheme.
For distance measurements, we decided to restrict application of the 
fundamental plane and SBF methods to types between -5 and -1 (E to L+) and of 
the Tully-Fisher method to types between 1 and 9 (Sa to Sm). Therefore, 7 
galaxies were excluded from the late-type sample (types -1, 0 and 10), and 2 
galaxies were excluded from the early-type sample (NGC~4440 and 
NGC~4531, both of type 1). The final sample thus contains 118 spiral, 
43 lenticular and 22 elliptical galaxies.

\subsection{Distances}

\subsubsection{Early-type galaxies}

   For early-type galaxies, we have used the G99 compilation of distances,
obtained from the fundamental plane method in $H$-band. According to these
authors, their accuracy should be about 21\%. These distances are based on an
assumed Virgo distance of 16~Mpc. To this compilation of 55 elliptical and 
lenticular galaxies, we added SBF distances from T01. It contains 35 
early-type galaxies in the Virgo cluster area, of which 25 are common with 
G99. Tonry's distances are calibrated independently of any assumed Virgo 
distance, and the mean shift compared to G99 amounts to:

\begin{equation}
\langle d_{\rm T} - d_{\rm G99} \rangle  =  0.64 \pm 0.84
\end{equation}

\noindent
after rejection of NGC~4638 ($d_{\rm G99} = 8.87$ Mpc and 
$d_{\rm T} = 21.68$ Mpc). As the mean difference of the two systems is not 
significant, we adopt an unweighted average of the two measurements as our 
distance measurement for this set of 25 early-type galaxies, and the 
uncorrected only available distance for the 40 other galaxies (30 from G99, 10 
from Tonry). Please note that both references include more uncertain 
measurements: in G99, they are identified as being due to velocity dispersions 
smaller than 100~km~s$^{-1}$; in T01, they appear in a separate table, named 
``table.poor''. We have kept these galaxies in our sample, but we generally 
exclude them from our statistical calculations.

\subsubsection{Late-type galaxies}

   For late-type galaxies, we also started from the G99 compilation (59
galaxies retained), obtained from the $H$-band Tully-Fisher method, with a 
claimed accuracy of about 16\%. We complemented this list with galaxies with 
$B$-band Tully-Fisher distances from E00 (41) and F98 (109). We do not use 
galaxies from these lists outside of 8 degrees from M~87 or with 
morphological types outside of our adopted range (1 - 9).

   To convert E00 distances into the G99 system, two corrections are done: 
first, we correct the distances to our adopted morphological type, in the 20 
cases where it differs from the LEDA type adopted in E00. Indeed, the 
intercept $b$ of the $B$-band Tully-Fisher relation derived by Theureau et al. 
(\cite{the97}) and adopted in E00 depends upon the morphological type. The
correction is given by:

\begin{equation}
d (T) = d_{\rm E00} \times 10^{0.2 \times (b(T_{\rm LEDA}) - b(T))}
\end{equation}

   Although the morphological type differences never exceed one unit, the 
correction can amount to 25\% in distance.

   Additionally, E00 distances are based on the Theureau et al. 
(\cite{the97}) calibration, which leads to 
$H_{\circ} \approx 55\,\rm km\,s^{-1}\,Mpc$; as E00 adopt a cosmic velocity 
of Virgo of $1200\,\rm km\,s^{-1}$, this implies a Virgo distance of 21.8~Mpc. 
Therefore, the conversion factor to the G99 assumed Virgo distance of 16~Mpc 
is:

\begin{equation}
d_{\rm E00}^{\rm cor} = 0.733\ d_{\rm E00}
\end{equation}

   To convert F98 distances into the G99 system, we simply apply the 
calibration correction derived from different adopted Virgo distances, namely 
21.5 Mpc and 16 Mpc:

\begin{equation}
d_{\rm F98}^{\rm cor} = 0.744\ d_{\rm F98}
\end{equation}

   As we have 65 galaxies with more than one distance estimate, we can make an 
intercomparison among the three sources of distance data, after their reduction
to the mean system. We find:

\begin{eqnarray}
\langle d_{\rm E00}^{\rm cor} - d_{\rm mean} \rangle & = & 1.21 \pm 0.29 \\
\langle d_{\rm F98}^{\rm cor} - d_{\rm mean} \rangle & = & -1.89 \pm 0.18 \\
\langle d_{\rm G99} - d_{\rm mean} \rangle & = & 1.13 \pm 0.25
\end{eqnarray}

where $d_{\rm mean}$ is the mean value of available distance measurements for a
given galaxy.

   All these shifts are significant. This means that our first guess of the 
conversion factors to the G99 system was not as successful as we could hope. 
We therefore repeated it with new conversion factors, until we reach on
average negligible shifts of the corrected systems to the mean one. This leads 
to the following adopted conversion factors, after rejection of two galaxies 
with discrepant distance measurements (NGC~4180 and NGC~4591):

\begin{eqnarray}
d_{\rm E00}^{\rm cor} & = & 0.678\ d_{\rm E00} \\
d_{\rm F98}^{\rm cor} & = & 0.832\ d_{\rm F98} \\
d_{\rm G99}^{\rm cor} & = & 0.930\ d_{\rm G99}
\end{eqnarray}

   We adopt these new conversion factors to put all three references into the
same system. Then, considering that the advantages of the $H$-band (low 
extinction correction) are compensated by the smaller slope of the Tully-Fisher
relation in $B$-band, we do not weigh each measurement and adopt a straight 
average of the available distance estimates for each galaxy. We will only use 
these mean distances to classify the galaxies among the various classes 
described in Sect.~6. We display in Fig.~\ref{histod} a histogram of these 
118 average distances. The mean value of 20~Mpc and the skewness of the 
histogram clearly show the background contamination of our sample. By 
comparison, the histogram of distances for early-type galaxies appears to be 
less contaminated by background galaxies (smaller mean distance and smaller 
skewness).

\begin{figure*}
   \begin{center}
   \scalebox{0.4}{\includegraphics{histod.eps}}
   \makebox[1.0cm]{}
   \scalebox{0.4}{\includegraphics{histodel.eps}}
   \end{center}
   \caption{Histogram of the adopted distances for the 118 spiral galaxies 
   (left) and the 65 early-type galaxies (right)}
   \label{histod}
\end{figure*}

\subsection{HI deficiency}

   As we are concerned with the distance determination of individual 
galaxies in the Virgo cluster, we quantify HI deficiency by means of a
distance-independent parameter DEF, based on the difference between the 
expected and observed logarithm of the mean (hybrid) HI surface density, 
$\Sigma_{\rm HI}$, that is

\begin{equation} \label{def1} 
\mbox{DEF}=\langle\log \Sigma_{\rm HI} (T^{\rm obs})\rangle-\log
\Sigma_{\rm HI}^{\rm obs}\;,
\end{equation}
with $\Sigma_{\rm HI}=F_{\rm HI}/a_{\rm opt}^2$, and where $F_{\rm HI}$ 
represents the corrected HI flux density integrated over the profile width in 
units of $\rm Jy\,km\,s^{-1}$ and $a_{\rm opt}$ represents the apparent optical
diameter in arcmin. For non-detections, an upper limit to HI content is 
estimated by assuming that the emission profile is rectangular with an 
amplitude 1.5 times the rms noise and width equal to that expected for a galaxy
of the same morphological type and luminosity, properly corrected for redshift 
broadening and viewing inclination (Solanes et al. \cite{sol96}; see also 
Haynes \& Giovanelli \cite{hay84}). The adopted standard values of 
$\langle\log\Sigma_{\rm HI}\rangle$ per type are: 0.24 for Sa, Sab; 0.38 for
Sb; 0.40 for Sbc; 0.34 for Sc; and 0.42 for later spiral types.

   Among our sample of 118 spiral galaxies, we have HI deficiency estimates for
106 galaxies, or 90\% of the sample.

   It has been suggested that Tully-Fisher distances may be underestimated
for highly HI deficient galaxies (T92; Fukugita et al. \cite{fuk93}). For this
reason, we have investigated possible differences in the rotation velocities of
HI deficient and HI normal galaxies in the 12 HI deficient clusters identified 
by Solanes et al. (\cite{sol01}), as well as the influence of the HI content on
the Tully-Fisher relationship of the galaxies in our sample. The results 
obtained allow us to conclude that HI deficiency does not affect our 
Tully-Fisher distances.

\section{Tolman-Bondi model}

   We use the Tolman-Bondi model of the Virgo cluster as defined in E99. Let 
us recall that the Tolman-Bondi model gives an analytical solution to 
Einstein's field equations for a spherically symmetric pressure-free density 
excess, embedded in an otherwise homogeneous universe. The parameters of the 
model are: the observed Virgo cluster velocity in the Local Group reference 
frame ($980\ \rm km\,s^{-1}$), the Virgocentric infall velocity of the Local 
Group ($220\ \rm km\,s^{-1}$), the Virgocentric density profile slope 
($\alpha = 2.85$), and the deceleration parameter of the background homogeneous
universe, taken as an Einstein - de Sitter universe for simplicity 
($q_{\circ} = 0.5$). Although there is a general agreement about the values of
3 of these parameters, the Virgocentric density profile slope is not well 
known, as it should reflect the distribution of the mass around the cluster
center, not only the galaxy (light) distribution. In E99, it has been 
constrained using 32 galaxies whose distances were known using their Cepheids, 
generally measured with the HST, through the PL-relation. For details about the
model, the adjustment of its parameters and the influence of changing these 
values, the reader is referred to T92, E99 and E00.

   For each galaxy, the angular distance to the Virgo center of mass (taken to 
be the position of M~87) completely defines the exact shape of the 
Tolman-Bondi S-curve in the velocity distance diagram, for a given choice of 
model parameters. Then, the observed recession velocity of the galaxy (in the 
Local Group reference frame) gives one to three possible distance ratios of the
galaxy to Virgo (see Fig.~\ref{ngc} for examples). The first and the third
values correspond to a galaxy falling into the cluster from in front or behind.
The second value corresponds to a true cluster member. Assuming that the 
Tolman-Bondi model can explain most of the observed positions of the galaxies
in the velocity-distance diagram means that we neglect other possible 
components, such as an expanding one suggested by T92, or galaxies projected
by chance onto the cluster, but taking part in the Hubble flow. Certainly,
not all galaxies in the Virgo cluster should follow the Tolman-Bondi model. For
instance, galaxies belonging to the virialized core of the cluster do not 
exhibit a velocity - distance relation. It is however difficult to disentangle 
those galaxies from our class 2, so that we include them into it. On the other 
hand, not all galaxies in the class 2 (defined as the descending branch of the 
S-shape Tolman-Bondi curve) belong to Virgo: for instance, at large angular
distances from M~87 the descending branch still exists, but is no longer 
related to proper Virgo galaxies. We assume that within 8 degrees of the 
cluster center, most of the galaxies on the descending branch are true cluster 
members. We will test this strong assumption in Sect.~6 by building a 
histogram of ``Virgo distances'', where each point is computed from a galaxy 
distance and its distance ratio to Virgo, as given by the Tolman-Bondi model. 

   To do this, we must select the most plausible one among the possible 
distance ratios for a given galaxy, when the model gives more than one value. 
Here, we made use of two criteria: we try to choose the value closest to the 
assumed Virgo distance (16~Mpc), and, for spiral galaxies, we take into account
the HI deficiency, assuming that a true member of the cluster has a higher 
probability to be deficient than a galaxy falling for the first time into the 
cluster. We therefore attribute to each galaxy a ``class number'', which is 1 
if we adopt the first value of the distance ratio to Virgo (galaxies falling 
from in front into the cluster), 2 if we adopt the second value (true cluster 
members), and 3 if we adopt the third value (galaxies falling from behind into 
the cluster).

\section{Distance to the Virgo cluster from Cepheids}

   There are 6 galaxies in or close to the Virgo cluster whose distance is 
known thanks to the HST observations in $V$ and $I$ bands, using the Cepheid 
period-luminosity relation. We use the distances published in Freedman et al. 
(\cite{fre01}) (without the uncertain metallicity correction). Five of the six 
distances cluster about a mean distance of 14.6~Mpc, while the sixth one 
(NGC~4639) gives a larger distance (21~Mpc). This is perfectly 
explained by the Tolman-Bondi model, and this was in fact the initial 
motivation of this study. Figure~\ref{ngc} gives the result of the application 
of the Tolman-Bondi model to these galaxies, with a dotted line showing the 
recession velocity of the galaxy, and an arrow showing the distance ratio to 
Virgo, for an adopted distance to Virgo of 15.4~Mpc (see below).
 
Two galaxies belong to the Virgo southern extension (NGC~4496A and NGC~4536) 
and lie at more than 8 degrees from M~87. However, their position in the 
velocity-distance diagram should be explained by our model, if we assume that 
no other cluster perturbes them. However, we find that the maximum velocity 
explained by our model for NGC~4536 is $1563\ \rm km\,s^{-1}$ at 
$0.83\ d_{\rm Virgo}$, too small compared to its observed velocity of 
$1641\ \rm km\,s^{-1}$. For this galaxy, our model predicts it to be in the 
class 3, with a distance of $1.79\ d_{\rm Virgo}$. This in turn would lead to 
an inacceptable Virgo distance of $d_{\rm Virgo} = 8.07\ \rm Mpc$, using the 
observed galaxy distance of 14.45~Mpc. In fact, it is well known that random 
velocities about 80 $\rm km\,s^{-1}$ exist for all galaxies and explain the 
velocity dispersion of small groups of galaxies (Gourgoulhon et al. 
\cite{gou92}). We therefore assume that this is the origin of the small 
discrepancy observed for NGC~4536.

Table~\ref{cephgal} gives the adopted distances to the six galaxies, the model
predicted distance ratios to the Virgo distance for the three possible 
solutions, the corresponding predicted ``Virgo distances'', the adopted class,
and the adopted Virgo distance for each galaxy. Now we discuss each galaxy in
some detail: for NGC~4548, only the second and third distances are 
plausible; but the large HI deficiency leads us to attribute this galaxy to 
class 2 as a true member of the Virgo cluster. Similarly, the large HI 
deficiency of NGC~4321 makes us attribute it to class 2. On the 
contrary, the insignificant deficiency of NGC~4639 leads to put it 
into class 3, with a derived Virgo distance intermediate between 
NGC~4321 and NGC~4548. For NGC~4535, the HI 
deficiency is low, but putting it into class 1 would lead to a larger derived 
Virgo distance than the previous galaxies, so we attribute it to class 2. 
NGC~4496A belongs to the Virgo southern extension and is not HI 
deficient; it is therefore tempting to put it into class 1, but again putting 
it into class 2 leads to a derived Virgo distance in better agreement with the 
remaining galaxies. Finally, NGC~4536 also belongs to the Virgo 
southern extension and has a small HI deficiency, and it may be put into class 
1 or 2, once its recession velocity is corrected for a random component. The 
resulting mean value of the six derived Virgo distances is 
$15.4 \pm 0.5\ \rm Mpc$, which we adopt as our estimate of the cluster 
distance.

At the referee's request, we have investigated what happens if a Virgo 
distance of 21.5~Mpc is preferred from independent arguments. Then, four of the
five galaxies classified in class 2 now fall into class 1 (infalling galaxies),
NGC~4639 is the only true member of the Virgo cluster among the six, and 
NGC~4548 position cannot be explained by the model. The high HI deficiency of 
NGC~4321 is contradictory. This alternative distance is clearly less probable 
according to these six galaxies with Cepheid distances. A definitive answer 
will only become available with a larger sample of Virgo galaxies with accurate
distances.

\begin{figure*}
   \begin{center}
   \resizebox{4.8cm}{4.8cm}{\includegraphics{NGC4321.eps}}
   \makebox[0.3cm]{}
   \resizebox{4.8cm}{4.8cm}{\includegraphics{NGC4496A.eps}}
   \makebox[0.3cm]{}
   \resizebox{4.8cm}{4.8cm}{\includegraphics{NGC4535.eps}}
   \raisebox{-1cm}{}
   \resizebox{4.8cm}{4.8cm}{\includegraphics{NGC4536.eps}}
   \makebox[0.3cm]{}
   \resizebox{4.8cm}{4.8cm}{\includegraphics{NGC4548.eps}}
   \makebox[0.3cm]{}
   \resizebox{4.8cm}{4.8cm}{\includegraphics{NGC4639.eps}}
   \end{center}
   \caption{Tolman-Bondi model for the six galaxies with Cepheid distances.
            The dotted line corresponds to the recession velocity of the 
            galaxy, while the arrows mark the expected distance ratio if 
            $R_{\rm VIR} = 15.4\ \rm Mpc$ (solid arrow) or
            $R_{\rm VIR} = 21.5\ \rm Mpc$ (dotted arrow)}
   \label{ngc}
\end{figure*}

\begin{table*}
   \parbox{11.5cm}{\caption{Adopted Cepheid distances to the 6 Virgo galaxies, 
   ratio to the Virgo distance given by the Tolman-Bondi model and 
   corresponding Virgo distances for each possible class, adopted class and 
   Virgo distance, and HI deficiency}}
   \begin{flushleft}
   \begin{tabular}{lccccccr}
   \hline\\[-0.3cm]
   Name & $d$ & 
   \multicolumn{3}{c}{$d/d_{\rm V}$} & 
   class & $d_{\rm V}$ & def\\ \cline{3-5}
        & Mpc & 
   $d_1$ & $d_2$ & $d_3$ &
         & Mpc & \\
   [0.3cm]\hline\\
   NGC~4321  & 14.32 &   0.68  &   0.95  &  1.71 &   2    &  15.07  & 0.49\\
             &       &  21.06  &  15.07  &  8.37 &        &      \\
   NGC~4496A & 14.52 &   0.75  &   0.92  &  1.76 & 2 (1)  &  15.78  & $-0.09$\\
             &       &  19.36  &  15.78  &  8.25 &        &      \\
   NGC~4535  & 14.79 &   0.77  &   0.96  &  1.91 & 2 (1)  &  15.41  & 0.19\\
             &       &  19.21  &  15.41  &  7.74 &        &      \\
   NGC~4536  & 14.45 &  (0.83) &  (0.83) &  1.79 & 1 or 2 & (17.41) & 0.25\\
             &       & (17.41) & (17.41) &  8.07 &        &      \\
   NGC~4548  & 15.00 &   0.23  &   1.07  &  1.30 &   2    &  14.02  & 0.83\\
             &       &  65.22  &  14.02  & 11.54 &        &      \\
   NGC~4639  & 20.99 &   0.47  &   1.03  &  1.45 &   3    &  14.48  & 0.10\\
             &       &  44.66  &  20.38  & 14.48 &        &      \\
   [0.3cm]\hline\\
   \label{cephgal}
   \end{tabular}
   \\[-3.5mm]
   \end{flushleft}
\end{table*}

\section{Mass of the Virgo cluster}

   Derivation of the Virgo cluster mass from the Tolman-Bondi model, and its 
comparison with the virial mass estimate, have been discussed in T92, E99 and
E00. Here, we give a summary of the useful formulae and derive the Virgo mass
for our adopted model parameters. Let $R_{\rm VIR}$ be the Virgo distance in
Mpc, $V_{\rm VIR}^{\rm cosm}$ the cosmic recession velocity of Virgo in 
$\rm km\,s^{-1}$ and $d$ the radius normalized to the Virgo distance. The mass
enclosed within $d$ is the product of the ``Einstein - de Sitter mass'' within 
the same radius by the mass excess due to the cluster. It is given in solar 
mass units by:

\begin{eqnarray}
M(d) & = & M(d)_{\rm EdS} \times (1 + k'\,d^{-\alpha})\\
M(d)_{\rm EdS} & = & 2.325\ 10^8\,q_{\circ}\,R_{\rm VIR}\,(V_{\rm VIR}^{\rm cosm})^2\,d^3
\end{eqnarray}

$k'$, the mass excess within $d = 1$, i.e. at the Local Group location, only
depends on the observed Virgo cluster velocity, the Virgocentric infall
velocity of the Local Group and the cosmological parameters of the adopted 
background universe ($q_{\circ}$ and $H_{\circ} \times t_{\circ}$). It does not
depend on the Virgocentric density profile slope $\alpha$. Its value is 
the same as in E99, namely $k' = 0.606$. For Virgo, we define the radius $d$ 
as corresponding to 8 degrees, which gives $d = 0.141$, in place of 0.105,
corresponding to 6 degrees in E99. 

With $V_{\rm VIR}^{\rm cosm} = 1200\ \rm km\,s^{-1}$ and 
$R_{\rm VIR} = 15.4\ \rm Mpc$, we get $M = 1.2\ 10^{15} M_{\sun}$. The virial 
mass of Virgo as given by Tully \& Shaya (\cite{tul84}) is:

\begin{equation}
M_{\rm virial} = 2.325\ 10^8 \times (\pi\,R_{\Omega}) \times (3\,\sigma_{\rm V}^2)
\end{equation}

\noindent
where notations are those from Tully \& Shaya. Their result transposed to 
15.4~Mpc gives $M_{\rm virial} = 6.9\ 10^{14} \ \rm M_{\sun}$, so that 
$M = 1.7\ M_{\rm virial}$. E99 found a coefficient 1.62 for their adopted
parameters of the Tolman-Bondi model.

   By comparison, B\"ohringer et al. (\cite{boh94}) estimate the mass of the 
M~87 sub-cluster from X-ray emission measured by ROSAT to 
$\sim (1.5 - 6) \times 10^{14} \ \rm M_{\sun}$ within a radius of 1.8~Mpc at
20~Mpc (5 degrees). Contributions from the M~49 and M~86 
sub-clusters are negligible, at about $(1 - 3) \times 10^{13} \ \rm M_{\sun}$. 
These values are confirmed by Schindler et al. (\cite{sch99}) who derive 
$2.1 \times 10^{14}\ \rm M_{\sun}$ within 1.5~Mpc around M~87 (at 
20~Mpc) and $0.87 \times 10^{14}\ \rm M_{\sun}$ within 0.75~Mpc around 
M~49. Our Tolman-Bondi mass for the same distance would be 
$M = 1.5\ 10^{15} \ \rm M_{\sun}$, almost one order of magnitude larger. We do 
not have any explanation to offer for this discrepancy, but we note that our 
large mass estimate and steep density profile are supported by Tully \& Shaya 
(\cite{tul98}), who find a Virgo mass of $M = 1.3\ 10^{15} \ \rm M_{\sun}$ 
from a modeling of the velocity field of the Local Supercluster, assuming
a mass-to-light ratio of $150 \rm M_{\sun} / \rm L_{\sun}$ in the field, but 
1000 for the Virgo cluster, and $\Omega_{\circ} = 0.3$.
\footnote{Although Tully \& Shaya do not specify their adopted distance to
Virgo, we assume it is the same as in Tully \& Shaya (\cite{tul84}), namely
16.8~Mpc.}

\section{Structure of the Virgo cluster}

   Disentangling the different components of the Virgo cluster is difficult and
somewhat subjective. However, the use of the Tolman-Bondi model allows us to
classify each galaxy into one of the three classes defined previously. We
did that independently for early-type and spiral galaxies. In each case, we 
have built several diagrams: the first one is the histogram of the ``Virgo 
derived distances'', and it is shown in Fig.~\ref{hisdc}. Both histograms now 
have a well-defined peak around the assumed Virgo distance and are symmetric, a
posteriori confirming our hypothesis. Mean distances in both cases are 
compatible with the assumed distance of 16~Mpc. However, the rms dispersion is 
slightly lower in the case of early-type galaxies: this may reveal that the 
spiral sample class attribution is still not fully satisfying, or that our 
restrictive hypothesis rejecting expanding components or projected galaxies in 
the Tolman-Bondi model are not completely fulfilled.

\begin{figure*}
   \begin{center}
   \scalebox{0.4}{\includegraphics{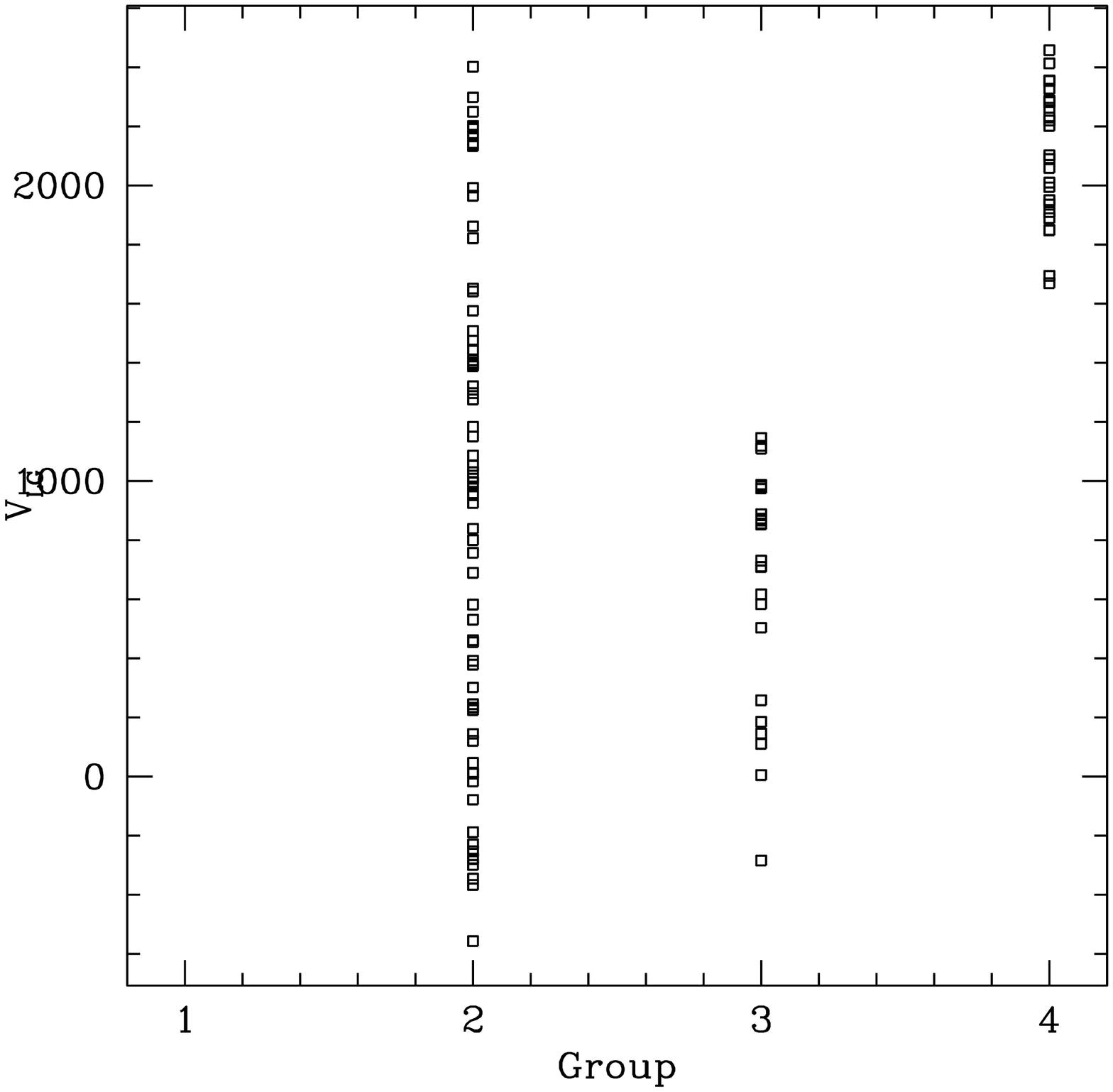}}
   \makebox[1.0cm]{}
   \scalebox{0.4}{\includegraphics{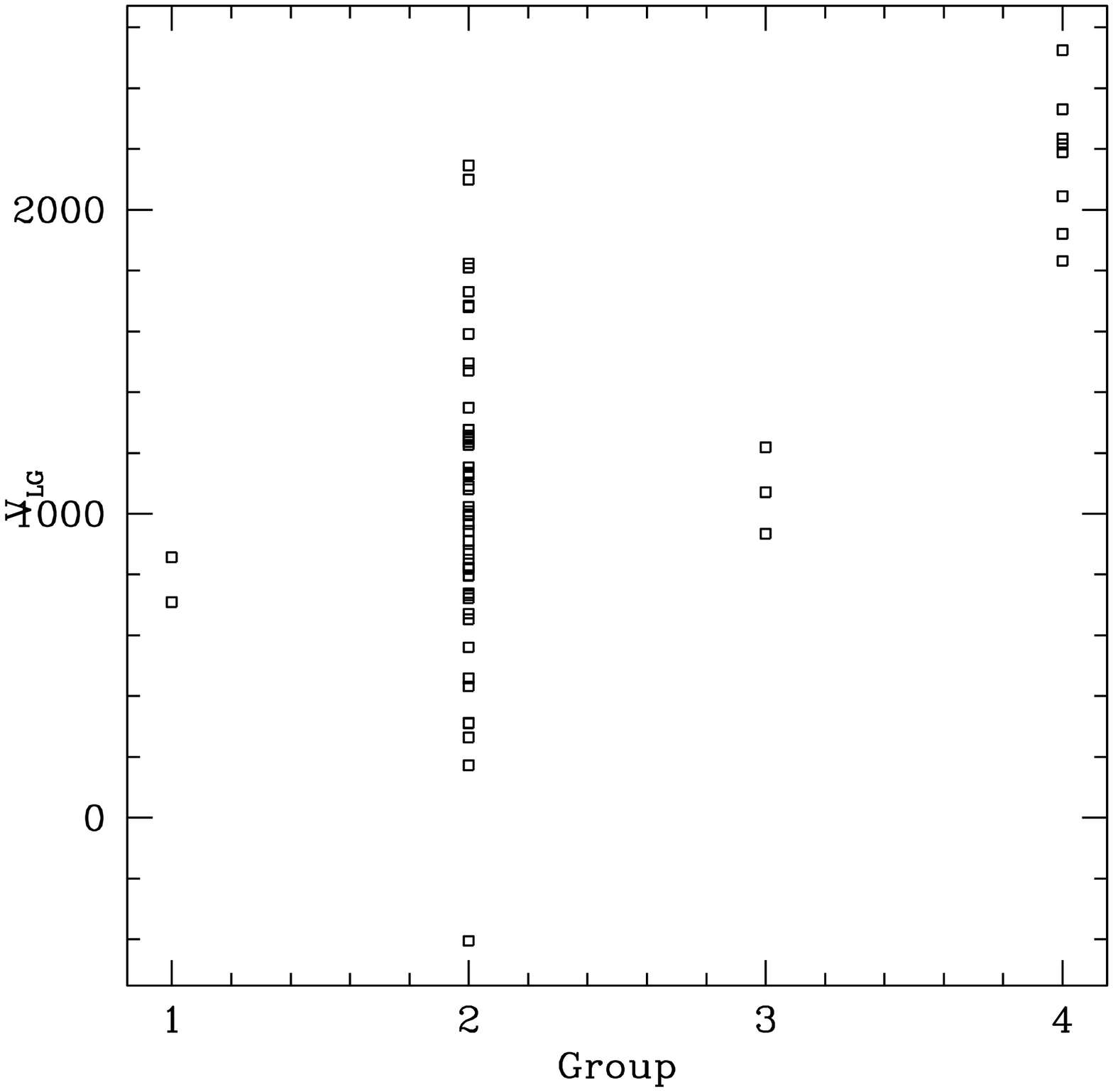}}
   \end{center}
   \caption{Distribution of the recession velocities among the different 
   classes for spiral galaxies (left) and early-type galaxies (right)}
   \label{velgrp}
\end{figure*}

   The second diagram gives the velocity repartition among the different 
classes, and is displayed in Fig.~\ref{velgrp}. In fact, we have split class 3 
into two classes, because it clearly contains high velocity objects (class 4,
above 1600 km s$^{-1}$), which are background galaxies, and low velocity 
objects (class 3, 0 -- 1300 km s$^{-1}$), which are infalling galaxies from
behind the cluster. We have not convincingly identified any spiral galaxy
belonging to class 1; the only two early-type objects classified in class 1 
exhibit low velocities (about 800 km s$^{-1}$), which is difficult to 
interpret; in fact, these two galaxies may be genuine members of the cluster 
(class 2) with bad distance measurements; indeed, both galaxies have a velocity
dispersion smaller than 100 km s$^{-1}$, which makes their fundamental plane 
distances quite uncertain.

\begin{figure}
   \begin{center}
   \scalebox{0.4}{\includegraphics{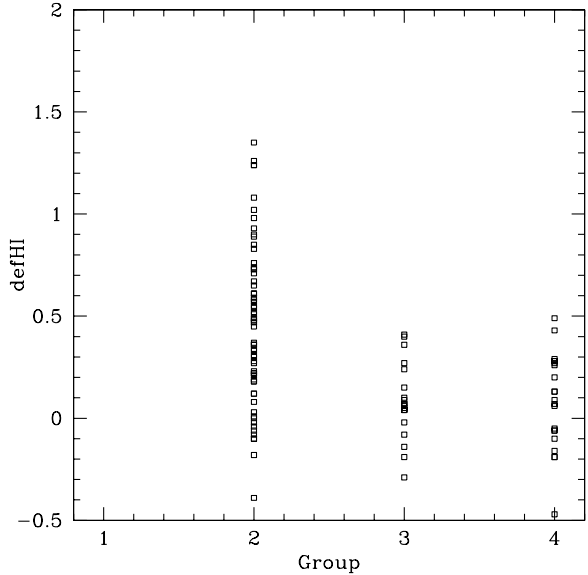}}
   \end{center}
   \caption{Distribution of HI deficiency among the different classes for 
   spiral galaxies}
   \label{defhi}
\end{figure}

   A third diagram gives the repartition of HI deficiency among the classes,
obviously only for spiral galaxies. It is displayed in Fig.~\ref{defhi}. It is
clear that there is no highly deficient galaxies among classes 1, 3 and 4, as
expected because we used HI deficiency as one of our classification criteria.

   Finally, Fig.~\ref{velcoo} displays the repartition on the sky of spirals
and early-type galaxies with different symbols for different classes. If we
compare the distribution of the two types of galaxies, it clearly appears that
the different classes concentrate into common regions. One can easily recognize
the W group around (12.3h, 6$^{\circ}$) and the M group around 
(12.2h, 13$^{\circ}$). Both groups have a mean distance ratio to Virgo of 2.1.
Galaxies at the same distance but slightly displaced from the main 
concentration have been classified as group halo. Finally, galaxies with 
intermediate distances between those of these groups and the Virgo distance, 
and classified into our class 3, are interpreted as belonging to extended
filaments extracted from the background group and falling into Virgo. Such is 
the case of the NGC~4222 group, which may link the M group to Virgo at
a mean distance ratio to Virgo of 1.2, and the NGC~4343 group, around 
(12.4h, 8$^{\circ}$), similar to the B group of G99 and introduced by de 
Vaucouleurs (\cite{dev61}) as its W' group, and which may link the W group to 
the Virgo cluster at a mean distance ratio to Virgo of 1.5. A small group 
around NGC~4639 at (12.7h, 14$^{\circ}$) can also be identified with the
eastern group of G99, although our group is less extended and less numerous; 
it could be linked to a remote group of galaxies around NGC~4746.

\begin{figure*}
   \begin{center}
   \scalebox{0.4}{\includegraphics{hisdc.eps}}
   \makebox[1.0cm]{}
   \scalebox{0.4}{\includegraphics{hisdcel.eps}}
   \end{center}
   \caption{Histogram of the ``Virgo derived distances'' for spiral galaxies
   (left) and early-type galaxies (right)}
   \label{hisdc}
\end{figure*}

\begin{figure*}
   \begin{center}
   \scalebox{0.4}{\includegraphics{velcoo.eps}}
   \makebox[1.0cm]{}
   \scalebox{0.4}{\includegraphics{velcooel.eps}}
   \end{center}
   \caption{Distribution on the sky of the spiral galaxies (left) and 
   early-type galaxies (right). Filled squares correspond to class 1, crosses 
   to class 2, open squares to class 3 and open triangles to class 4}
   \label{velcoo}
\end{figure*}

In Table~\ref{groups}, we propose a new nomenclature for all these groups, 
based on the name of its main galaxy, together with a list of the members we 
identified in this study. These lists are obviously not complete, because of 
the lack of distance measurements of some other well-known members. 
Table~\ref{others} lists other galaxies which have not been attributed to these
groups nor to the main cluster, and are classified as background (class 4),
infalling from behind (class 3), or from in front (class 1).

\begin{table*}
   \parbox{17.5cm}{\caption{List of groups outside of the main Virgo cluster, 
   with their mean distance ratio to Virgo and a list of members with known 
   distance}}
   \begin{flushleft}
   \begin{tabular}{lcll}
   \hline\\[-0.3cm]
   Name & $d/d_{\rm V}$ & Location & Members \\
   [0.3cm]\hline\\
   NGC~4168 & 2.13 & M group  & NGC~4168, NGC~4189, NGC~4193, NGC~4200, IC~769, IC~3061, IC~3099 \\
            & 2.07 & M halo   & NGC~4067, NGC~4152, IC~3074 \\
   NGC~4222 & 1.24 & M infall & NGC~4222, IC~3033, IC~3066, IC~3105, UGC~7249 \\
   NGC~4261 & 2.11 & W group  & NGC~4180, NGC~4197, NGC~4215, NGC~4233, NGC~4235, NGC~4259, NGC~4260, \\
            &      &          & NGC~4261, NGC~4273, NGC~4281, IC~3225, CGCG~42-1, CGCG~42-36 \\
            & 2.17 & W halo   & IC~776, UGC~7579 \\
   NGC~4343 & 1.46 & W infall & NGC~4252, NGC~4316, NGC~4318, NGC~4343, NGC~4353, NGC~4376, NGC~4390, \\
            &      &          & NGC~4411A, NGC~4434, NGC~4451, NGC~4466, IC~3115, IC~3322A, UGC~7423 \\
   NGC~4639 & 1.41 &          & NGC~4620, NGC~4633, NGC~4639, NGC~4651, IC~3742 \\
   NGC~4746 & 1.92 &          & NGC~4746, UGC~8085, UGC~8114 \\
   [0.3cm]\hline\\
   \label{groups}
   \end{tabular}
   \\[-3.5mm]
   \end{flushleft}
\end{table*}

\begin{table}
   \parbox{5cm}{\caption{List of galaxies outside the main Virgo cluster 
   and not attributed to any group, with their distance ratio to Virgo}}
   \begin{flushleft}
   \begin{tabular}{lcc}
   \hline\\[-0.3cm]
   Name & $d/d_{\rm V}$ & Class \\
   [0.3cm]\hline\\
   NGC~4417 & 0.39 & 1 \\
   NGC~4488 & 0.46 & 1 \\
   NGC~4591 & 2.21 & 4 \\
   NGC~4598 & 1.92 & 4 \\
   NGC~4698 & 1.44 & 3 \\
   IC~3298  & 2.27 & 4 \\
   IC~3483  & 1.22 & 3 \\
   UGC~7697 & 2.34 & 4 \\
   [0.3cm]\hline\\
   \label{others}
   \end{tabular}
   \\[-3.5mm]
   \end{flushleft}
\end{table}

   After excluding all the galaxies which do not belong to the main cluster 
according to our classification, we are left with 71 spiral and 52 early-type 
galaxies. 67\% of the original sample with distances are therefore considered
as true members of the cluster; not surprisingly, the spiral sample exhibits a 
larger contamination (40\%) than the early-type sample (20\%). Among our 5 
references, the proportion of true members varies from 56\% (E00) and 61\% 
(F98) to 83\% (T01).

\section{Comparison of Virgo distances}

   Now that we have classified all the galaxies of our sample into the 
different classes, we can return to the different distances adopted by 
the various authors to the Virgo cluster.

   We will first derive an average ratio of distances from a given reference to
the predicted distance given by the Tolman-Bondi model and an adopted Virgo 
distance of 15.4~Mpc. For the $i^{\rm th}$ galaxy measured by reference $j$, we
have:

\begin{eqnarray}
r_{ij} & = & \frac{d_{ij}}{(d/d_{\rm V})_i \times 15.4} \\
r_j & = & \langle r_{ij} \rangle
\end{eqnarray}

   In the second test we restrict our sample to true Virgo galaxies 
(class 2) and compute the mean distance for each reference. The results will
depend much less on our adopted Tolman-Bondi model but should confirm it.

   Table~4 gives the results. We can see that for G99 and 
T01, the agreement between the resulting mean Virgo distances from both 
methods (Col.~3 and Col.~4, respectively) and the value adopted by these 
authors (Col.~5) is satisfying. On the contrary, for F98 the resulting mean 
Virgo distances are consistent from both methods, but smaller than their 
adopted value: this is certainly due to inclusion of background galaxies into 
their sample; indeed, among the 49 galaxies of their ``fiducial sample'', 14 
are classified by us as non-members of the Virgo cluster. For E00 the result 
is inverse and the mean error is large: it seems therefore that there is a 
discrepancy between their adopted Hubble constant (from Theureau et al., 
\cite{the97}) and the Virgo distance we derive from their data. 

For these two references, we adopt as the mean Virgo distance the average over
the two determinations (Col.~3 and Col.~4). This gives 18.9 and 23.2~Mpc for
F98 and E00, respectively. If we apply to these distances the conversion 
factors adopted in Sect.~2.3.2 to reduce F98 and E00 to a mean Virgo distance
of 16~Mpc, namely 0.832 and 0.678 respectively, we now get values close to 
16~Mpc (15.7 and 15.8~Mpc, respectively).

\begin{table*}
   \parbox{13.5cm}{\caption{Mean ratio of galaxy distances to Tolman-Bondi 
    predictions for $d_{\rm Virgo} = 15.4$ Mpc, resulting mean Virgo distance, 
    comparison with the mean distance restricted to true Virgo galaxies and to 
    the distance adopted by each of our 5 references}}
   \begin{flushleft}
   \begin{tabular}{lcccccc}
   \hline\\[-0.3cm]
   Reference (Galaxy type) & Ratio & $d_{\rm V}$ & $d_{\rm V}^{\rm true}$ 
   (N, $\sigma$) & $d_{\rm V}^{\rm ref}$ \\
   [0.3cm]\hline\\
   G99 (E)    & $1.017 \pm 0.031$ & $15.7 \pm 0.5$ & $15.3 \pm 0.5$ (36, 2.94) & 16.0 \\
   T01 (E) & $1.084 \pm 0.024$ & $16.7 \pm 0.4$ & $17.0 \pm 0.4$ (29, 1.99) & $17.0 \pm 0.3$ \\
   G99 (S)    & $1.025 \pm 0.025$ & $15.8 \pm 0.4$ & $15.7 \pm 0.5$ (41, 3.10) & 16.0 \\
   F98 (S)      & $1.221 \pm 0.027$ & $18.8 \pm 0.4$ & $19.1 \pm 0.5$ (67, 4.27) & $21.5 \pm 2.3$ \\
   E00 (S)    & $1.487 \pm 0.053$ & $22.9 \pm 0.8$ & $23.6 \pm 1.1$ (23, 5.51) & 21.8 \\
   [0.3cm]\hline\\
   \label{ratios}
   \end{tabular}
   \\[-3.5mm]
   \end{flushleft}
\end{table*}

   However, it is important to keep in mind that the clean samples of true
members may still be affected by the incompleteness bias (Fouqu\'e et al.
\cite{fou90} for the specific case of the Virgo cluster; Teerikorpi
\cite{tee97} for a general discussion), which may lead to average cluster 
distances which are too short. To estimate the amount of this bias, we 
have calibrated once more the B-band Tully-Fisher relation using the 21 
calibrators from Freedman et al. (\cite{fre01}), and the 51 true members of 
Virgo according to our classification, which best match the calibrator 
properties (thus excluding peculiar, interacting, HI-truncated galaxies, and 
restricting to morphological types between 2 and 8, inclinations between 37 and
90 degrees, $\log V_{\rm max}$ larger than 1.7), with data extracted from the 
LEDA database. As usual, the slope of the Tully-Fisher relation is determined 
from the Virgo sample (more numerous), and the intercept from the calibrators. 
The rms dispersion is similar for both samples (0.41 for calibrators, 0.53 for 
Virgo sample). The direct fit (appropriate to distance determinations) gives:

\begin{equation}
M_{\rm B} = -6.28 (\pm 0.39) \log V_{\rm max} - 6.31 (\pm 0.09)
\end{equation}

The intercept for the Virgo sample, $24.96 \pm 0.11$ gives a distance modulus 
of $31.27 \pm 0.14$, or a distance of $18.0 \pm 1.2$ Mpc, which is larger than
our adopted distance from Cepheids. If we divide the sample of 51 Virgo 
galaxies equally into large and small $\log V_{\rm max}$, we get mean distances
of:

\begin{eqnarray}
\mu & = & 31.24 \pm 0.09, {\rm for} \log V_{\rm max} > 1.992 \\
\mu & = & 31.28 \pm 0.12, {\rm for} \log V_{\rm max} \le 1.992
\end{eqnarray}

It can therefore safely be concluded that the incompleteness bias does 
not affect significantly our sample of true Virgo members (it would lead to a 
smaller mean distance for galaxies with small rotation velocities, which are 
more biased).

It can still be argued that the bias may be hidden, because the slope of 
the Tully-Fisher relation, when determined from the Virgo cluster sample, may 
be biased due to a possible incompleteness of the sample at small 
$\log V_{\rm max}$. This is however not observed, as the direct slope 
determined from the calibrators is $-5.24 \pm 0.66$, while an incompleteness 
bias would predict a steeper slope as compared to the slope determined from the
Virgo sample ($-6.28 \pm 0.39$).

The 0.3 mag difference between the Cepheid and Tully-Fisher distances is 
worrying. Tenants of the long distance scale will see evidence that the 
Tully-Fisher distance is the correct one, while most galaxies with Cepheid 
distances are infalling on the front side of the cluster.

\section{Conclusions}

   In this paper, we have investigated how the relativistic Tolman-Bondi model
as applied in E99 gives constraints on the Virgo cluster mass and distance, and
allows one to disentangle its quite intricate structure. Distances to
183 Virgo galaxies from 5 references have been used and averaged, together with
HI deficiency parameters for spirals, to classify the galaxies into 4 different
distance classes: an infalling component in front of Virgo, the Virgo cluster 
itself, an infalling component behind the main cluster, and background groups. 
The main results of the present study are: 

\begin{itemize}
\item Among the 6 galaxies in the Virgo area with Cepheid distance measures,
NGC~4639 is well-known to be discrepant. Our model nicely explains 
this apparent discrepancy, and the resulting mean distance to Virgo using all 6
calibrators is $15.4 \pm 0.5$ Mpc. However, the mean Tully-Fisher distance
derived from 51 spiral galaxies classified as true members of the cluster in 
the present study, and calibrated using the same Cepheid distance system 
(Freedman et al. \cite{fre01}) is larger ($18.0 \pm 1.2$ Mpc).
\item The mass of the Virgo cluster derived from our model is large: 
$M = 1.2\ 10^{15} M_{\sun}$, which corresponds to 1.7 virial mass. Our adopted 
mass profile is steeper than the light distribution (anti-biasing).
\item Apart from the well-known background groups which once contaminated Virgo
samples, we have identified a number of galaxies at intermediate distances 
between these groups and the cluster itself, which we interpret as filaments
extracted from the background groups and falling into the cluster from behind.
\item We have not been able to securely identify galaxies in front of the 
cluster and falling into it. This supports Sandage \& Tammann (\cite{san76}) 
who always argued that apparently foreground galaxies in the Virgo cluster 
direction were in fact cluster members.
\item Distances to the Virgo cluster adopted by F98 and E00 differ from what 
would be expected using the distances predicted by our model and the mean
ratio of published to predicted distances. They also differ from the mean of 
the published distances restricted to the true Virgo members. We also note that
the distances published in E00 are highly dispersed, even for true Virgo 
members.
\end{itemize}

\begin{acknowledgements}
   We have made use of the LEDA database (http://leda.univ-lyon1.fr), 
supplied by the LEDA team at the CRAL - Observatoire de Lyon (France). We
warmly thank all the LEDA team members for their effort. We also thank Riccardo
Giovanelli and Martha Haynes for making their Arecibo General Catalog, from
which we have extracted the HI data used in this study, available to us. 
Finally, we wish to thank the referee, Pekka Teerikorpi, for his very 
constructive comments. T.S. acknowledges support from a fellowship of the 
Ministerio de Educaci\'on, Cultura y Deporte of Spain. C.B. aknowledges ESO for
a visiting position in Santiago during which this work was started.
\end{acknowledgements}

\end{document}